\documentstyle[11pt]{article}

\columnwidth\textwidth

 \def\RR{{\bf{R}}} \def\ZZ{{\bf{Z}}}
\def\D{\,\mbox{\rm d}\!\!\;} \def\d{\partial}

\begin{document}

\begin{center}
\vbox to 2cm {}
CHIRAL CHARGED FERMIONS, ONE DIMENSIONAL QUANTUM FIELD THEORY AND VERTEX 
ALGEBRAS\\[5mm]

Florin Constantinescu\\ Fachbereich Mathematik, Johann Wolfgang
Goethe--Universit\"at Frankfurt, Robert--Mayer--Strasse 10, D--60054
Frankfurt am Main, Germany\\[5mm]

G\"unter Scharf\\Institut f\"ur Theoretische Physik, Universit\"at
Z\"urich\\Winterthurerstr.~190 CH--8057 Z\"urich, Switzerland\\[20mm]
\end{center}

{\bf Abstract}

We give an explicit $L^2$--representation of chiral charged fermions
using the Hardy--Lebesgue octant decomposition. In the " pure"
case such a representation was already used by M.~Sato in holonomic
field theory. We study both "pure" and " mixed" cases. In the compact
case we rigorously define unsmeared chiral charged fermion operators
inside the unit circle. Using chiral 
fermions we orient our findings towards a
functional analytic study of vertex algebras as one dimensional quantum
field theory.
\newpage

\section{Introduction}

Let $\RR^n=\cup_{\alpha=1}^{2^n}\Gamma_\alpha$ be the octant
decomposition of $\RR^n$ by octants $\Gamma_\alpha$ considered with
their parity. Each $L^2$--function $\psi(x)$, $x\in\RR^n$ has a
decomposition
\begin{equation}
\psi(x)=\sum^{2^n}_{\alpha=1}\psi_\alpha(x+i\Gamma_\alpha0),i=\sqrt{-1}
\end{equation}
where $\psi_\alpha$ are $L^2$--boundary values in tubes
$\RR^n+i\Gamma_\alpha$ induced by the octant decomposition in the
conjugate Fourier variable. Remark that two different boundary values
$\psi_\alpha,\varphi_\beta$, $\alpha\not=\beta$ of $\psi,\varphi\in
L^2(\RR^n)$ are orthogonal, i.e. $\int\psi_\alpha(x)\varphi_\beta(x)\D x=0$.
We call (1) Hardy--Lebesgue octant decomposition. Certainly, a compact
version in which $\RR$ is replaced by the unit circle $S^1$ is also
possible by using the Cauchy indicatrix. In this case we have the
classical Hardy decomposition. By means of this decomposition we
introduce in section 2 chiral charged fermions by an explicit
representation in the antisymmetric Fock space ${\cal F}(L^2(\RR))$. This
representation parallels usual canonical anticommutation relations. 
Charged Fermions polarize the antisymmetric Fock space. Accordingly 
a graduation of the form
\begin{equation}
{\cal F}(L^2(\RR))=\bigoplus_{n=0}^\infty{\cal F}_n(L^2(\RR)=
\bigoplus_{n=0}^\infty\bigoplus_{n_1+n_2=n}{\cal F}_{n_1,n_2}(L^2(\RR)
=\bigoplus_{n_1,n_2\ge0}{\cal F}_{n_1,n_2}(L^2(\RR))
\end{equation}
appears where $n_1,n_2\ge0$ refer to the octant decomposition and
collect all $\Gamma_\alpha$ with $n_1$ positive and $n_2$ negative
components. The case
$n_1=n_2=0$ corresponds to the vacuum. We call spaces ${\cal F}_{n_1,0}$,
${\cal F}_{0,n_2}$ " pure" and others " mixed. The "
pure" fermionic sectors were already used by M.~Sato and coworkers
[1] in holonomic field theory. We insist here on the "mixed"
sectors. In the compact case we rigorously define analytically
continued chiral, charged fermionic operators inside the unit circle.

Our findings are analyzed from the point of view of vertex operator
algebras realized as one--dimensional quantum field theory. There is no
spectral condition in momentum space. Translation invariance and
especially locality play the central role. An interesting point is that
our quantum fields appear as boundary values of analytic (operator-
valued) functions, a property which in the standard Wightman quantum
field theory is reserved to vacuum expectation values being a
consequence of the spectral condition (and Poincar\'e invariance).  In
section 3 in order to establish full contact to vertex algebras we have
to pay a price by giving up complex conjugation. This loss is fully
compensated by rigorous operator product expansion. Formal algebraic
machinery of vertex algebras is enriched by functional analytic
framework. The example of chiral charged fermions and other examples 
worked out in this
paper can be extended to more general vertex algebras.

\section{Fock space representations of charged fermions}

In order to explain what we are going to do let us consider the
well--known representation of canonical anticommutation relations in
the antisymmetric Fock space ${\cal F}(L^2(\RR))$:
\begin{equation}
\Big(A(f)\psi\Big)^n(x_1,\ldots,x_n)=(n+1)^{1/2}\int
  dx\overline{f(x)}\psi^{n+1}(x,x_1,\ldots,x_n)
\end{equation}
\begin{displaymath}
\Big(A^*(f)\psi\Big)^n(x_1,\ldots,x_n)=n^{-1/2}\sum^n_{i=1}(-1)^{i-1}f(x_i)
\end{displaymath}
\begin{equation}
\times\psi^{n-1}(x_1,\ldots,\hat{x}_i,\ldots,x_n)
\end{equation}
and in the unsmeared form:
\begin{equation}
\Big(A(x)\psi\Big)^n(x_1,\ldots,x_n)=(n+1)^{1/2}\psi^{n+1}(x,x_1,\ldots,x_n)
\end{equation}
\begin{displaymath}
\Big(A^*(x)\psi\Big)^n(x_1,\ldots,x_n)=n^{-1/2}\sum^n_{i=1}(-1)^{i-1} 
\delta(x-x_i)
\end{displaymath}
\begin{equation}
\times\psi^{n-1}(x_1,\ldots,\hat{x}_i,\ldots x_n)
\end{equation}
Here $\psi=(\psi^0,\psi^1,\ldots,\psi^n,\ldots)$ where
$\psi^n=\psi^n(x_1,x_2,\ldots,x_n)\in L^2(\RR^n)$ is an element of the
antisymmetric Fock space, $\psi^0$ is the vacuum and the hat denotes the
missing variable. On the vacuum we have
\[
A(f)\psi^0=0,\quad f\in L^2(\RR).
\]
The smeared $ A(f)$ and $A^*(f)$ are densely defined, closed operators
with $A^*(f)=A(f)^*$ being the adjoint of $A(f)$. The unsmeared $A(x)$ is an
operator whereas $A^*(x)$ is only a bilinear form. The maps
\[
f\mapsto A(f),\quad f\mapsto A^*(f)
\]
are antilinear and linear, respectively, and 
\begin{equation}
A(f)=\int dx\overline{f(x)}A(x),\quad A^*(f)=\int
dxf(x)A^*(x).
\end{equation}
The operators $A(f)$ and $A^*(f)$ satisfy the canonical anticommutation
 relations
\begin{eqnarray}
\lefteqn{\{A(f),A(g)\}=\{A^*(f),A^*(g)\}=0}\\
&&\{A(f),A^*(g)\}=(f,g)
\end{eqnarray}
where $(f,g)$ is the scalar product in $L^2(\RR)$. The operators
$A(f)$, $A^*(f)$ are bounded with norm
\begin{equation}
\|A(f)\|=\|A^*(f)\|=\|f\|_2
\end{equation}
Now we introduce charged fermions $a(f),a^*(f), f\in L^2(\RR)$ by the
following defining relations in antisymmetric Fock space:
\begin{displaymath}
\Big(a(f)\psi\Big)^n(x_1,\ldots,x_n)=(n+1)^{1/2}\int\overline{f_+(x)}
\psi^{n+1}(x,x_1,\ldots,x_n)\D x
\end{displaymath}
\begin{equation}
+n^{-1/2}\sum^n_{j=1}(-1)^{j-1}f_-(x_j)\psi^{n-1}(x_1,\ldots,\hat 
{x}_j,\ldots,x_n)\end{equation}
\begin{displaymath}
(a^*(f)\psi)^n(x_1,\ldots,x_n)=(n+1)^{1/2}\int\overline{f_-(x)} 
\psi^{n+1}(x,x_1,\ldots,x_n)\D x
\end{displaymath}
\begin{equation}
+n^{-1/2}\sum^n_{j=1}(-1)^{j-1}f_+(x_j)\psi^{n-1}(x_1,\ldots,\hat{x}_j,\ldots,x_n)
\end{equation}
where $f=f_++f_-$ is the Hardy--Lebesgue decomposition of $f$.

In compact form the relations (11) and (12) can be given with the help 
of $A^{\#}$, ($A^{\#}$ stays for $A$ or $A^*$) as
\begin{eqnarray}
&&a(f)=A(f_+)+A^*(f_-)\\
&&a^*(f)=A(f_-)+A^*(f_+)
\end{eqnarray}
where $A^*(f_{\pm})=A(f_{\pm})^*$.
Remark the similarity of (13) and (14) to two dimensional
(time--independent) Dirac fermions which are also defined in the
(antisymmetric) Fock space over the direct sum $H_+\oplus H_-$. The
difference is that for Dirac fermions we start in momentum space and
both $H_+$ and $H_-$ are copies of the same $L^2$ Hilbert--space. In
addition the natural conjugation in $H_+$ and $H_-$ accounts for the
charge conjugation, see for instance [2]. Writing in this case
$f=(f_+,f_-)$, $f_{\pm}\in H_{\pm}\equiv L^2(\RR)$ the formal difference
is that for Dirac fermions the second term in $a(f)$ has
$\overline{f_-}$ instead of $f_-$, $a^*(f)$ being the adjoint of $a(f)$.
In this way no mixed linear/antilinear dependence on $f$ appear, which
is typical for (13) and (14), see later. On the other hand it is well 
known that putting
together chiral fermions of opposite chirality one obtains the massless
Dirac fermion.

Time--zero Dirac fermion operators satisfy CAR relations whereas chiral
charged fermions $a(f),a^*(f)$ defined above satisfy anticommunitation
relations of the form 
\begin{eqnarray}
\{a(f),a(g)\}&=&\{a^*(f), a^*(g)\}=0\\
\{a(f),a^*(g)\}&=&\langle f,g\rangle
\end{eqnarray}
where in contradistinction to (9) $\langle f,g\rangle$ is no longer the
scalar product in $L^2(\RR)$ but it is
\begin{equation}
\langle f,g\rangle=\int\overline{f_+(x)}g_+(x)\D
x+\int\overline{g_-(x)}f_-(x)\D x.
\end{equation}
We call the attention of the reader to the fact that in our case it is
not possible to take $\overline{f_-}$ instead of $f_-$ in (13) as in the
case of Dirac fermions, which would turn the inner product
$\langle\cdot,\cdot\rangle$ into the usual scalar product
$$(f,g)=\int\overline{f_+(x)}g_+(x)\,dx+\int\overline{f_-(x)}g_-(x)\,dx
=\int\overline{f(x)}g(x)\,dx,$$
because the relations (15) are no longer true ((16) remains valid).
In spite of not being a scalar product, $\langle\cdot,\cdot\rangle$ is
positive definite. Indeed
\begin{displaymath}
  \langle
  f,f\rangle=\int\overline{f_+(x)}f_+(x)+\int\overline{f_-(x)}f_-(x)\D x 
\end{displaymath}
\begin{equation}  
=(f,f)=\|f\|^2_2
\end{equation}
because $\int\overline{f_+(x)}g_-(x)\D x
=\int\overline{f_-(x)}g_+(x)\D x=0$ for arbitrary $f, g\in L^2(\RR)$.

As remarked in the introduction the antisymmetric Fock space decomposes according to
\begin{equation}
{\cal{F}}(L^2(\RR))=\bigoplus_{n_1,n_2\ge0}{\cal{F}}_{n_1,n_2}=
\bigoplus_{n=0}^\infty\bigoplus_{n_1+n_2=n}{\cal F}_{n_1,n_2}=
\bigoplus^\infty_{n=-\infty}\bigoplus_{n_1-n_2=n}{\cal{F}}_{n_1,n_2}
\end{equation}
giving rise to " pure" and " mixed" states already
mentioned. Here $n_1,n_2$ are the numbers of the $a^*$ and $a$ operators, 
respectively, applied to the vacuum.

Neither $a(f)$ nor $a^*(f)$ annihilates the vacuum as in the neutral
case. Vacuum expectation values of $a^{\#}(f)$ satisfy neutrality
condition (i.e. they do vanish if the number of $a$--operators is not equal
to the number of $a^*$--operators) and can be given in closed form. They are
Gram determinants in the pure cases and a kind of generalized Gram "
determinants" in the mixed case. Moreover they are bounded
operators:
\begin{equation}
\|a^*(f)\|=\|f\|_2.
\end{equation}
It is interesting to remark that the proof of (20) doesn't follow
directly from CAR because $\langle\cdot,\cdot\rangle$ in (16) is not a
scalar product. Nevertheless the reader can check without difficulty
that the usual boundedness proof for fermions [3] can be easily adapted
because it only uses positivity of $\langle\cdot,\cdot\rangle$. 

The explicit Fock space realization of chiral charged fermions (11) 
 and (12) can be taken over to the compact case $S^1$. 

It is somewhat unpleasant that in the non--compact as well as in the
compact case $a(f)$ and $a^*(f)$ do not show sharp linear or
anti--linear dependence on $f$ such that the unsmeared $a(z)$, $a^*(z)$, 
$z\in\RR$ or $S^1$ cannot be looked at as operator-valued 
"kernels" for $a(f)$, $a^*(f)$. In the compact case this point will
be discussed (and improved) in the next section. This is the main point
of this paper.

Coming to the end of this section let us remark that the notation $a(f)$
for the chiral fermion is not fortunate. Indeed we want to look at
$a(f)$ as a one dimensional field operator and as such a better notation
would be $\Phi(f)$. The reason we choose $a$ instead of $\Phi$ is
explained in the next section where the unsmeared $a(z)$ will be
interpreted as a member of a vertex algebra.

\section{Towards vertex algebras as one--dimensional quantum field
  theories}

Vertex algebras have been related to the two--dimensional Wightman
theory [4]. The results of the preceeding section together with results
obtained in [5] show that a natural variant is to start with a
one--dimensional quantum field theory. Certainly we cannot expect
bona fide Wightman fields but the fields we are suggesting have better
behaviours. 

Indeed, in Wightman theory, as a consequence of spectral condition together
with Poincar\'e invariance, vacuum expectation values show up as certain
boundary values of analytic functions. The fields by themselves do not
appear as boundary values.

In contradistinction to that, our one--dimensional fields are boundary
values of operator-valued analytic functions. But in order to achieve full
contact to vertex algebras we still have to do some work which we
accomplish in the frame of our example of chiral fermions. Some other
examples follow later. In order to start let us remark that strong 
tools in quantum field theory and string theory like operator product 
expansion (OPE) have to be formulated in the unsmeared form, which is
from a functional analytic point of view not well defined. In conformal 
quantum field theory a rigorous way out are
vertex algebras in which OPE is formulated in the frame of formal power
series, formal Laurent expansions and formal distribution theory.

What we claim in this paper is that the algebraic framework of vertex
algebras can be realized in functional analysis and this implies the
search for a kernel calculus. We decide to give up complex conjugation
and to define all our fields as operators in a given complex domain which will be the
unit disc $|z|<1$. Even in the case of chiral fermions it is not at all
clear that this is possible. The following explicit representation of
$a(z)$ and $b(z)$ (instead of $a^*(z)$) as operators inspired from the
formal unsmeared version of (11) and (12) solves the problem:
\begin{displaymath}
\Big(a(z)\psi\Big)^n(z_1,\ldots,z_n)=(n+1)^{1/2}\psi_+^{n+1}\Big(z^{-1},z_1,\ldots,z_n\Big)\\
\end{displaymath}
\begin{equation}
+n^{-1/2}\sum^n_{j=1}(-1)^j\frac1{z-z_j}\psi^{n-1}(z_1,\ldots,\hat{z}_j,\ldots,z_n)\\
\end{equation}
\begin{displaymath}
\Big(b(z)\psi\Big)^n(z_1,\ldots,z_n)=(n+1)^{1/2}\psi_-^{n+1}(z,z_1,\ldots,z_n)\\
\end{displaymath}
\begin{equation}
+n^{-1/2}\sum^n_{j=1}(-1)^j\frac1{z-z^{-1}_j}\psi^{n-1}(z_1,\ldots,
\hat{z}_j,\ldots,z_n),
\end{equation}
where $z_j\in S^1$, $|z|<1$. We are now going to show that $a(z), b(z)$
are densely defined operators in fermionic Fock space with a domain of
definition presented below. In some sense complex conjugation
$z\to\overline z$ is simulated in (21) and (22) by the inversion $z\to z^{-1}$.
Although these relations are similar to the unsmeared version of (11)
and (12), they show particularities not present in (11), (12) which will
be made clear by progressing in this section.

We use the notations $\psi_{\pm}^{n+1}(z,z_1,\ldots,z_n)$ in order to
denote Hardy components of $\psi^{n+1}$ in the first variable. For
instance for the pure case
$$
\psi(z_1,\ldots,z_n)=\det(\varphi^i(z_j))_{i,j=1,\ldots,n},\ \
\varphi^i(z)\in L^2(S^1)
$$
we have
$$
\psi_{\pm}(z_1,\ldots,z_n)=\left|\begin{array}{l}\varphi^1_{\pm}(z_1)\ldots
    \varphi^n_{\pm}(z_1)\\ \varphi^1(z_2)\ldots \varphi^n(z_2)\\
    \varphi^1(z_n)\ldots\varphi^n(z_n)\end{array}\right|.
$$
Further $\psi^n_+(z^{-1},\ldots)$ means " wrong boundary
value" i.e. we first take $\psi_+^n(z\ldots)$ and then replace $z$ by
$z^{-1}$. In order to make sense of this type of boundary value we will
restrict $\psi^n$ to Laurent polynomials instead of Laurent series.
This restriction enables us to rigorously define $a(z)$ and $b(z)$ as
Fock space operators. Indeed, we identify ${\cal F}(L^2(S^1))$ with the
corresponding space of analytic functions inside and outside the unit
circle obtained by the Hardy decomposition (which are Laurent series)
and define $a(z), b(z)$ through (21), (22) on the dense domain in
${\cal F}(L^2(S^1))$ obtained by cutting the Laurent series to Laurent
polynomials. In order to define products of such operators
(beside introducing the normal ordering) we have to extend the
definition to some
semi--infinite Laurent series. This will be discussed later on in this
section.

The reader can check anticommutation relations of $a$ and $b$, in
particular
\begin{equation}
\{a(z),b(w)\}=\delta(z-w)
\end{equation}
where now
$$\{a(z),b(w)\}=a(z)b(w)|_{|z|>|w|}+b(w)a(z)|_{|w|>|z|}=$$
$$={1\over z-w}\Bigl\vert_{|z|>|w|}+{1\over w-z}\Bigl\vert_{|w|>|z|}
\eqno(24)$$
and $\delta(z-w)$ replaces the formal $\delta$--function [4]:
$$\delta(z-w)=z^{-1}\sum_{n\in\ZZ}\Big(\frac{w}{z}\Big)^n=w^{-1} 
\sum_{n\in\ZZ}\Big(\frac{z}{w}\Big)^n=\delta(w-z).\eqno(25)$$
Some hints are given below.
This is exactly the state of affair in the vertex algebra frame where by 
now $a(z)$, $b(z)$ are genuine densely defined operators in Fock space
for $|z|<1$.

Some remarks are in order concerning (23). Let us introduce first some
notations similar to those in section 2:
$$a(z)=a_1(z)+a_2(z)\eqno(26)$$
$$b(z)=b_1(z)+b_2(z)\eqno(27)$$
where
$$\Big(a_1(z)\psi\Big)^n(z_1,\ldots,z_n)=(n+1)^{1/2}\psi_+^{n+1} 
\Big(z^{-1},z_1,\ldots,z_n\Big)\eqno(28)$$
$$\Big(a_2(z)\psi\Big)^n(z_1,\ldots,z_n)=n^{-1/2}\sum^n_{j=1}(-1)^j\frac1{z-z_j}
$$
$$\times\psi^{n-1}(z_1,\ldots,\hat{z}_j,\ldots,z_n)
\eqno(29)$$
and similar relations for $b_1(z)$ and $b_2(z)$. The operators 
$a_i(z),b_i(z)$, $i=1,2,|z|<1$ are defined on Laurent polynomials
$\psi^n$ but in order to write down (24) we need products of $a_i$ and
$b_i$. Such products are a priori not defined. Let us discuss the case 
$a_1(z)b_2(w)$ with $|z|>|w|$. Here we have to extend
the domain of definition of $a_1(z)$ from Laurent polynomials to some
semi--infinite Laurent series. Indeed, for the first critical term
$-\frac1{w-z^{-1}_1}$ in $b_2(w)$ we have 
\begin{eqnarray*}
&&a_1(z)\left(-\frac1{w-z_1^{-1}}\right)=a_1(z)\left(\frac{z_1}{1-wz_1}\right)\\
&=&
a_1(z)\Big(\sum^\infty_{n=0}w^nz_1^{n+1}\Big)=\sum^\infty_{n=0}w^na_1(z)(z_1^{n+1})\\
&=&\sum^\infty_{n=0}w^nz^{-n-1}=\frac1{z-w}
\end{eqnarray*}
for $1>|z|>|w|$ and this is the correct result. The other term
$\frac1{w-z}b_1(z)$ with $1>|w|>|z|$ results from $b_1(w)a_2(z)$
after extending the domain of definition of $b_1(z)$. We have 
used $|z|,|w|<|z_1|,|z_1|^{-1}$.

The reader is asked to check the usual axioms of vertex algebras for our 
Fock space representation (21, 22). In particular the translation operator
$T$ (which coincides with the Virasoro generator $L_{-1}$) and the
state field correspondence is induced by $a(z),b(z)$ as strongly
generating set of fields [4] chapter 4 together with relations
$$a(z)|0>|_{z=0}=-z_1^{-1},\quad b(z)|0>|_{z=0}=z_1
\eqno(30)$$
where $|0>$ is the vacuum and $z_1,z_1^{-1}$ appear as Laurent
monomials interpreted as states in our Fock space. Here the vacuum is
defined to be the constant Laurent polynomial. Derivatives and
Wick products of $a(z),b(z)$ which are defined algebraically in vertex
algebras allow straightforward interpretation as Fock space operators. 
The operators $a(z), b(z)$ defined in (21, 22) also satisfy
$$\{a(z), a(w)\}=0=\{b(z), b(w)\}
\eqno(31)$$
independent of the relative position of $z$ and $w$ inside the unit
circle. We give some details.

The proof of (31) is similar to the proof of (23) and is based on the
equation
$${1\over z-z_1}\Bigl\vert_+=0$$
where + means "wrong boundary value" as defined above. Ordering
conditions on $z$ and $w$ are not necessary here as it should be on
symmetry reasons. We mention that the relations (28), (29) can be
expressed by using the Cauchy kernel in order to generate $\psi_\pm$
from $\psi$.

The translation operator $T$ can be identified as
$$T\psi^n(z_1,\ldots ,z_n)=\sum_{i=1}^n{\d\over\d z_i}\psi^n(z_1,
\ldots ,z_n)\eqno(32)$$
and satisfies [4]
$$[T, Y(a,z)]=\d Y(a,z)\eqno(33)$$
where $Y(a,z)$ is the general element of the vertex algebra generated by
$a(z)$ and $b(z)$ and $a$ is the Fock space vector
$$Y(a,z)|0\rangle |_{z=0}=a.$$
In particular, using (30) we have
$$Y(-z_1^{-1},z)=a(z),\quad Y(z_1,z)=b(z).\eqno(34)$$
A more precise definition of $Y(a,z)$ will be given below after
introducing the Wick product and the smearing out operation on $a(z)$
and $b(z)$.
The Wick product $:a(z)b(z):$ is defined as
$$:a(z)b(z):\>=a_2(z)b(z)+b(z)a_1(z)\eqno(35)$$
and it is again a densely defined operator for $|z|<1$. From the
explicit relations (21), (22) it is clear why $a_1(z)$ cannot stand in
front of $b_2(z)$ (and $b_1(z)$ in front of $a_2(z)$). Formulas giving
the Wick product through contour integrals which are well known in
conformal quantum field theory and string theory are in our frame
rigorous equalities between densely defined operators, instead of being
used formally as usual.

Let us remark that the relations (26), (27) suggest the
interpretation of $a(z)$ and $b(z)$ as densely defined operator-valued
analytic functionals (hyperfunctions). This property persists in all
other examples of vertex algebras to follow and is very natural if one
remembers that elements of vertex algebras are usually defined as formal
Laurent series with operator coefficients. It suggests a one-dimensional
Wightman quantum field theory (cf. section 4). In the particular case of
chiral charged fermions the densely defined smeared operators $a(f),
b(f)$, $f\in L^2({\bf R})$ can be shown to be bounded, as this was
the case with chiral fermions in section 2 but we refrain from giving
details because this property is incidental here and is not true for
other examples to follow. Instead let us discuss the nature of the
singularity in $a(z), b(z)$ when $z$ passes from the interior to the
boundary of the unit circle. This problem is related to the operator
product expansion which was claimed to be under rigorous functional
analytic control. To see this the reader can write down for
$z=re^{i\theta}, r<1$
$$a(f)=\lim_{r\to 1-}{1\over 2\pi}\int\limits_0^{2\pi}a(z)f(e^{i\theta})
e^{i\theta}d\theta\eqno(36)$$
$$b(f)=\lim_{r\to 1-}{1\over 2\pi}\int\limits_0^{2\pi}b(z)f(e^{i\theta})
e^{i\theta}d\theta\eqno(37)$$
where $a(z), b(z)$ are explicitly given by (21), (22) and $f\in L^2(S^1)$.
Formally but suggestive we write instead of (36), (37):
$$a(f)={1\over 2\pi i}\int a(z)f(z)\,dz\eqno(38)$$
$$b(f)={1\over 2\pi i}\int b(z)f(z)\,dz\eqno(39)$$
with integrals on the unit circle. These operator relations are
understood as applied to Laurent polynomials or even on semi-infinite
Laurent  series according to the definition of the operators $a(z),
b(z), |z|<1$ and the extension of their domain of definition given
above. The complex conjugation of test functions which was essential for
the linear/anti-linear nature of (11), (12) was here completely ignored.
This might spoil some special operator properties (like boundedness)
but, as remarked above, this is not the point. What counts is a rigorous
natural (linear) interpretation of $a(z), b(z)$ as operator boundary
values (operator-valued hyperfunctions). This interpretation is
particularly rewarding when one introduces operator product expansions.
Indeed the rigorous definition of OPE takes place inside the unit circle
and segregates the expected singularity. It is in perfect agreement with
formal physical work in string and conformal field theory. As an example
we can look at products of the form $a(z)b(w)$ or $b(z)a(w)$ with
$|z|>|w|$ which can be analysed exactly as above in the context of
verifying (23). This is in contrast to ordinary quantum field theory
where even in the free case the corresponding definition looks rather
heavy. Indeed in order to formulate a rigorous OPE (and define Wick
products) of free fields one first writes down formal expressions and
only in a second step gives a smeared out definition over the Fourier
transform (see for instance [6]). 

Specializing in (38), (39)
to monomial test functions we obtain operators satisfying
anticommutation relations. They have interesting representations in our
Fock space which can be traced back to the reproducing property of the
Cauchy kernel. We use them to give a direct definition of $Y(a,z)$ which
also works for the general case of vertex algebras by constructing first
the Fock space vector $a$ to which $Y(a,z)$ is associated; $a\to
Y(a,z)$. Let us start with the formal expression
$$a(z)=\sum_{n\in\ZZ}a_nz^{-n-1}\eqno(40)$$
common in vertex algebras considered as formal Laurent series with
operator coefficients. In our setting the coefficients $a_n$ are
$$a_n=a(z^n),\quad n\in\ZZ\eqno(41)$$
where (41) is the particular case of (38) with $f(z)=z^n$. By a similar
formula we define the coefficients $b_n, n\in\ZZ$. It is interesting to
remark that both $a_n$ and $b_n, n\in\ZZ$ are not only densely defined
as was the case with $a(z), b(z)$, but in addition the set of Laurent
polynomials is an invariant domain of definition with respect to forming
products (this was not the case with $a(z), b(w)$; remember the
necessity of the argument ordering). Now in order to make a long story
short, in the general case we have to consider monomial smearing (41) of
the entire generating set of the vertex algebra under consideration [4].
Arbitrary products of such operators with $n<0$ applied to the vacuum
generate the Fock vector $a$ and the correspondence $a\to Y(a,z)$ is
given by the formula (4.4.5) of [4] by means of Wick products.
  
Let us finally remark that the operator properties of $a(z)$,
$b(z)$ are similar to those of vertex operators $V_{\pm1}(z)$ and
their smeared out counterparts obtained in [5], which were introduced by
different methods in a different framework.
Indeed both $V_{\pm1}(z)$ exist as densely defined operators for $|z|<1$
but do not have a meaning for $|z|\ge1$. As far as the smeared versions
of $a,b$ (and $V_{\pm1}$) is concerned [5] we are allowed to approach the
unit circle from the interior. We want to stress that the equations (21, 22)
show in a clear way
that the smeared operators $a(f)$, $b(f)$, $f\in L^2(S^1)$ involve both
$f_+$ and $f_-$ from the Hardy decomposition $f=f_++f_-$ in spite of the 
fact that $a(z)$ and $b(z)$ are defined only inside the unit
circle. This is a central point of our approach. Certainly the
similarity between $a(z)$ and $b(z)$ on one side and $V_{+1}(z)$,
$V_{-1}(z)$ on the other side is a consequence of the correspondence
between bosons and fermions in two dimensions. It appears here in
conjunction with [5] at the true operator level.

After the example of chiral fermionic vertex algebra as one-dimensional
quantum field theory we proceed to other examples. The simplest one is
the $\tilde u(1)$ theory generated by currents
$$J(z)=a(z)b(z)\eqno(42)$$ 
where we left out the Wick dots. The precise
relations giving $J(z)$ as densely defined operator in Fock space can be
assembled from (21) and (22). It is clear that the typical square of the
Cauchy kernel makes its appearance. The central statement is the
locality of $J(z)$:
$$\{J(z), J(w)\}=\delta'_z(z-w)=-\delta'_w(z-w)\eqno(43)$$ 
and can be
verifield in the sense of operators by a direct computation. It also
follows by twice applying the Dong lemma (cf. section 4) to $a(z)$,
$b(z)$ and $J(z)$. Other examples include non-Abelian generalisation of
$\tilde u(1)$ like current algebras with currents (see for instance [7])
$$J^a(z)=\sum_{i,j=1}^Na_i(z)t^a_{ij}b_j(z)\eqno(44)$$
where $a_i, b_j$ are independent chiral fermions and $t_{ij}^a$ are
transformation matrices in the defining representation of $su(N)$ such
that
$${\rm Tr}t^at^b=\delta_{ab}$$
$$\sum_at^a_{ij}t^b_{kl}=\delta_{il}\delta_{jk}-{1\over N}\delta_{ij}
\delta_{kl}$$
$$[t^a,t^b]=\sum_cf_{abc}t^c$$
$$\sum_{a,b}f_{abc}f_{abd}=2N\delta_{cd}$$
with $f_{abc}$ being the structure constants. The (mutual) locality of
the currents $J^a(z)$ can again be verified by direct computation.

Finally we remark that we can obtain from the charged (complex fermions
(21), (22) explicit representations of real fermions which in turn can
be used for generating explicit representations of $\widehat{so}(N)$
current algebras at level one or even higher levels (see [7]).

\section{Remarks and conclusions}
The aim of this paper is twofold. First we gave an explicit
representation of chiral charged fermions in Fock space insisting on
what we called " mixed states". Second we modified the above
(unsmeared) representation in order to get full contact to 
definitions, methods and techniques in vertex algebras [4]. This is the
main point of the paper. Although we have presented only
simple examples it is clear that the present results can be generalized.
Summarizing we have shown how functional analysis penetrates vertex
algebras formulated as one-dimensional quantum field theory.

It would be interesting to develope our findings into a general axiomatic
approach to vertex algebras. In the above mentioned framework of
one-dimensional quantum field theory one should start with a locally
convex algebra of distributional test functions on the circle
(Borchers algebra) on which vertex algebra elements are defined as
(unsmeared) operators inside the unit disc. If positivity is expected
then the factorization and Hilbert space completion common in
Wightman theory boil down to a symmetrization property of the Borchers
algebra consistent with symmetry properties of the given vacuum
expectation values. The Cauchy indicatrix and its variants present in
our examples discussed above has to be
replaced by some reproducing kernels characterizing the given vertex
algebra. This remembers proposals in [8] section 5 and [9]
section 8. 

We want to stress the very appealing idea of a one-dimensional quantum
field theory with fields being operator boundary values (operator-valued
hyperfuctions), as opposed to the standard Wightman theory where this
property is reserved to vacuum expectation values. This might have
drastic consequences. For example we mention the Dong lemma of vertex
algebra which in our framework turns out to be a triviality following
from the trasitivity of (mutual) locality via weak locality (i.e.
commutativity inside expectation values). It is a simple example of the
celebrated Borchers transitivity of local quantum field theory.

Another remark concerns the quality of our operators representing
elements of vertex algebras. They are densely defined but might be ugly
for instance concerning closability (cf. [5]). But nobody would expect
more from them as long as they can be multiplied.
The examples presented in this paper, i.e. chiral charged fermions and
current algebras are easily understood in this framework. In fact, the
program can be extended to vertex algebras with a fermionic
representation. We do not know how to obtain explicit representations
(if any) of the type (21, 22) for general lattice vertex algebras,
although our feeling is that such representations could exist.

Last but not least an explicit operator realization of the
type (21), (22) and its smeared counterpart can be used to study the
$C^*$--content of vertex algebras. This seems to be usefull in the
context of recent developement at the interface between strings and
non--commutative geometry; see for instance [10].

{\bf Acknowledgement.} We thank A.~Hoffmann for discussions on the
subject of functional analytic study of vertex algebras.

\vskip 0.5cm
{\bf References}
\vskip 0.5cm
\noindent
1. M.~Sato, T.~Miwa, M.~Jimbo, Aspects of Holonomic Quantum Fields, 

in Complex analysis, microlocal calculus and relativistic quantum

theory, D.~Iagolnitzer ed., Lecture notes in physics, {\bf 126},
Springer, 1980

\noindent
2. A.\,L.~Carey, S.N.M.~Ruijsenaars, Acta Appl.\,Math. {\bf 10}
(1987),1

\noindent
3. H.~Araki, W.~Wyss, Helv.Phys.Acta {\bf 37} (1964), 136

\noindent
4. V.~Kac, Vertex algebras for beginners, second edition,
University 

lecture series, vol {\bf 10}, Providence, RI:
Am.Math.Soc.,1998

\noindent
5. F.~Constantinescu, G.~Scharf, Commun.\,Math.\,Phys. {\bf 200}
(1999), 275

\noindent
6. A.S.~Wightman, L.~G{\aa}rding, Arkiv f\"or Fysik {\bf 28} (1964), 129

\noindent
7. P. di Francesco, P.~Mathieu, D.~Senechal, Conformal Field Theory,

Springer 1997

\noindent
8. M.\,R.~Gaberdiel, P.~Goddard, Axiomatic conformal field theory,

hep--th/9810019

\noindent
9. R.\,E.~Borcherds, Vertex algebras, q--alg/9706008

\noindent
10. F.~Lizzi, Noncommutative Geometry, Strings and Duality, 

hep-th/9906122

\end{document}